\documentclass[twocolumn,showpacs,aps]{revtex4}
\usepackage{epsfig}

\begin{document}
\title{Real-space Renormalization Group Approach for the Corner Hamiltonian}
\author{Kouichi Okunishi}
\affiliation{Department of Physics, Niigata University, Igarashi 2, Niigata 950-2181, Japan.}
%\inst{Department of Physics, Niigata University, Igarashi 2, Niigata 950-2181, Japan}
%\address{Department of Physics, Niigata University, Igarashi 2, Niigata 950-2181, Japan.}
\date{\today}
 
\begin{abstract}
%\abst{
We present a real-space renormalization group approach for the corner Hamiltonian, which is relevant to the reduced density matrix in the density matrix renormalization group.
A set of self-consistent equations that the renormalized Hamiltonian should satisfy in the thermodynamic limit is also derived from the fixed point of the recursion relation for the corner Hamiltonian.
We demonstrate the renormalization group algorithm for the $S=1/2$ XXZ spin chain and show that the results are consistent with the exact solution.
We further examine the renormalization group for the $S=1$ Heisenberg spin chain and then discuss the nature of the eigenvalue spectrum of the corner Hamiltonian for the nonintegrable model.
%}
\end{abstract}

\pacs{75.10.Jm, 05.50.+q}
%\kword{corner Hamiltonian, CTM, DMRG, real-space renormalization group}

%%%%%%%%%%%%%%%%%%%%%%%%

\maketitle

%%%%% introduction %%%%%%
\section{Introduction}

Since Wilson's pioneering work for the Kondo impurity problem,\cite{wilson,KWW} the numerical renormalization group(NRG) procedure has been developed  extensively for the various low-dimensional quantum many-body systems.
In particular, the density matrix renormalization group(DMRG) enables us to obtain  reliable results for the ground states of the one-dimensional(1D) quantum many-body systems.\cite{White,springer,Scholl}
However, the DMRG has an essential difference from Wilson's approach;
In the Wilson's NRG, the lower-energy states of the Hamiltonian are selected to become relevant in low-energy physics, while the larger eigenvalue states  of the reduced density matrix of the ground state are more important in the DMRG.
Although the variational principle for the matrix product form of the ground state wavefunction lies behind the DMRG\cite{or}, what kind of the low energy theory is generated by the DMRG is not discussed from the Wilson's renormalization group view. 
Of course, it may be difficult to answer such a fundamental question for general cases.  
In this paper, we want to present a possible approach to address the question.

In order to discuss the reduced density matrix in the DMRG, it is essential to bring  Baxter's corner transfer matrix(CTM), which is defined as the partition function of a quadrant for the 2D classical lattice systems.\cite{Bax1,Bax2,Baxbook}
The peculiar feature of the CTM for the integrable model is that the CTM $A$ can be exponentiated as 
\begin{equation}
A(\mu) \propto e^{-\mu K},
\label{ctm-k}
\end{equation}
where $\mu$ is a spectral parameter and $K$ is the corner Hamiltonian(the detailed definition is given by (\ref{defctmh})).\cite{Bax1,Bax2,Baxbook,thacker}
As pointed out in ref.\cite{ctmrg}, the reduced density matrix in the DMRG is related to a product form of the CTMs.
Then the reduced density matrix $\rho$ for the integrable model in the thermodynamic limit can be also written as
\begin{equation}
\rho\propto e^{-\alpha K},
\label{rho-k}
\end{equation}
where $\alpha$ is a certain parameter.
On the basis of eq. (\ref{rho-k}), we can view that the DMRG provides a low-energy effective theory of the corner Hamiltonian, suggesting that the real-space renormalization group approach also works for the corner Hamiltonian.
Although eqs. (\ref{ctm-k}) or (\ref{rho-k}) are established only for the integrable models,  we may extract implications about the role of the reduced density matrix for general cases from the corner Hamiltonian. 

In addition,   the reduced density matrix in the DMRG  is recently attracting much attention in the context of quantum information theory.\cite{holzhey,vidal,Calabrese}
When  a reduced density matrix $\rho$ is defined for a subsystem with a cut  of  certain geometry, the entanglement entropy $S= - {\rm tr}(\rho \ln \rho$) plays an important role in  understanding how quantum states are entangled through the other subsystem.
The reduced density matrix or CTM in the DMRG is interestingly associated with the entanglement entropy of the 1D quantum systems with the half-infinite cut,
implying that the property of the corner Hamiltonian is theoretically  important from the quantum information theory  point of view.
For instance, Peschel has studied a generalization of the corner Hamiltonian for the various free fermion models, which provides  new insight in the studies of the properties of the quantum states.\cite{chung,peschel,peschel2}

In this paper, we discuss the real-space renormalization group approach for the corner Hamiltonian.
In the next section, we formulate the recursion relation  and the iterative diagonalization algorithm for the corner Hamiltonian. 
We then derive the self-consistent equations for the renormalized corner Hamiltonian in the thermodynamic limit.
We also rewrite the self-consistent equation so as to be more relevant to the block Hamiltonian in the DMRG formulation.
In \S 3, we demonstrate the corner Hamiltonian NRG for  the XXZ spin chain. 
We find that the resulting low-energy eigenvalue spectrum of the corner Hamiltonian  reproduces well the exact solution for the massive case.
For the XXZ chain in the massless region, we analyze the size dependence of the NRG spectrum and verify the logarithmic size dependence expected by the conformal field theory(CFT)\cite{peschel-troung,cardy}.
In \S 4, we further apply the NRG to the $S=1$ Heisenberg spin chain, which is a typical example of the nonintegrable spin model.
We then find that  the dominant structure of the eigenvalue spectrum of the corner Hamiltonian can be scaled by one parameter to the corresponding spectrum(logarithm) of the reduced density matrix.
Moreover, we analyze the ``spectral flow'' of the corner Hamiltonian for the $S=1$ bilinear-biquadratic chain.
In \S 5, we summarize  our results and discuss further implications of the eigenvalue spectrum of the corner Hamiltonian.

\section{Formulation}

\subsection{definitions}
In this section we consider the  $S=1/2$ XXZ spin chain for convenience. 
However, the arguments in the following can be generalized straightforwardly to the general 1D quantum systems.
We write the local Hamiltonian of the XXZ spin chain as
\begin{equation}
 h_{n,n+1}= S_n^xS_{n+1}^x+ S_n^yS_{n+1}^y+\Delta S_n^zS_{n+1}^z,
\label{xxzh}
\end{equation}
where $\vec{S}$ is the $S=1/2$ spin(not Pauli) matrix.
The matrix element of  (\ref{xxzh}) is labeled by the spin indices $s_n$, $s_{n+1}$ and  $s'_n$, $s'_{n+1}$.
However, we do not show  explicitly such spin indices without necessity.

Let us denote the Hamiltonian of N spins as
\begin{equation}
H_{\rm N}=\sum_{n=1}^{N-1}  h_{n,n+1} ,
\end{equation}
for which the free boundary condition is basically assumed.
In the context of the DMRG, $H_{\rm N}$ is the right-half(or left-half) block of the total Hamiltonian and, in this labeling of the site index, $n=1$(N) is assigned to the center(edge) of half of the system. 

\begin{figure}[ht]
\begin{center}
\epsfig{file=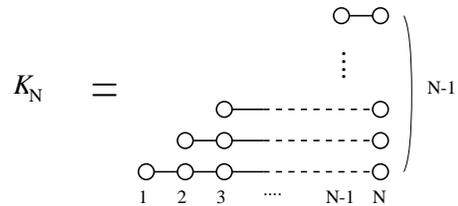,width=6cm}
\end{center}
\caption{Graphical representation of  corner Hamiltonian $K_{\rm N}$. 
White circles indicate spins and the lines connecting two spins mean the local interaction $h_{n,n+1}$. The horizontal lines are ``stacked'' to become the corner Hamiltonian.}
\label{tag-fig1}
\end{figure}

We further define the corner Hamiltonian as
\begin{equation}
K_{\rm N}=\sum_{n=1}^{N-1} n h_{n,n+1},
\label{defctmh}
\end{equation}
whose graphical representation is depicted in Fig. \ref{tag-fig1}.
As noted in the introduction, the corner Hamiltonian is the generator of the CTM  for the integrable models. Thus, (\ref{defctmh}) contains $N(N-1)/2$ local bonds, corresponding to the quadrant of the 6-vertex model.
Here, we note that $K_{\rm N}$ is clearly an Hermite matrix.

\subsection{recursion relation}

In order to formulate the real-space renormalization group for the corner Hamiltonian, we have to set up the recursion relation for matrices having different dimension sizes.
For this purpose, we introduce some notations for a $2^{\rm N}\times2^{\rm N}$ matrix $X$:
\begin{equation}
X_{\rm N}^* = \delta(s_1,s'_1) X_{\rm N},
\end{equation}
where the row index of $X_{\rm N}$ is labeled $s_2\cdots s_{\rm N+1}$ and thus $X_{\rm N}^*$ has the index $s_1,s_2\cdots s_{\rm N+1}$.  
Similarly, we also use $X_{\rm N-1}^{**} = \delta(s_1,s'_1)\delta(s_2,s'_2) X_{\rm N-1}$.

\begin{figure}[ht]
\begin{center}
\epsfig{file=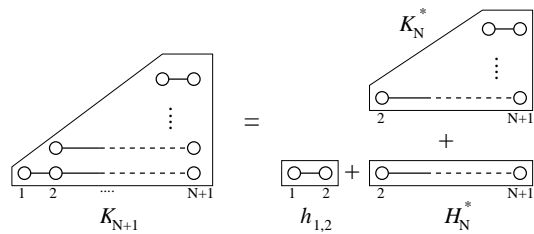,width=7cm}
\end{center}
\caption{Graphical representation of recursion relation (\ref{recursion1}).}\label{tag-fig2}
\end{figure}

We then construct the recursion relation of the corner Hamiltonians between N and N+1.  As is illustrated in Fig.\ref{tag-fig2}, we can decompose $K_{\rm N+1}$ into three pieces and then find
\begin{equation}
K_{\rm N+1}=  h_{1,2} + H_{\rm N}^* + K_{\rm N}^*.
\label{recursion1}
\end{equation}
However, this relation contains both of $H$ and $K$, which is not convenient for capturing the eigenvalue structure of the corner Hamiltonian directly.
In order to eliminate  $H^*_{\rm N}$ in (\ref{recursion1}), we exploit a supplemental recursion relation for $ H^*_{\rm N}$ which is represented as 
\begin{equation}
H^*_{\rm N} =K^*_{\rm N}-K^{**}_{\rm N-1}.
\label{recursionh}
\end{equation}
We can thus construct the recursion relation consisting of the corner Hamiltonians
\begin{equation}
K_{\rm N+1}=   h_{1,2}+2K^*_{\rm N} - K^{**}_{\rm N-1}.
\label{recursionk1}
\end{equation}
A key point on (\ref{recursionk1}) is that we have derived the recursion relation between the corner Hamiltonians for three sizes (N+1,N,N$-$1) rather than for the two sizes (N+1,N) by eliminating  $H^*_{\rm N}$.
Here, it should be noted that such a construction of the recursion relation is almost parallel to the ``logarithm'' of  Baxter's recursion relation for the CTMs\cite{Baxbook}.

We next convert the bases of the matrices into the representation  diagonalizing $K_{\rm N}$:
\begin{equation}
K_{\rm N} U_{\rm N} = U_{\rm N} \omega_{\rm N},
\end{equation}
where $ \omega_{\rm N}$ is a diagonal matrix of the eigenvalues of $K_{\rm N}$, and $U_{\rm N}$ is a unitary matrix consisting of the corresponding eigenvectors. 
We assume that the elements of $ \omega_{\rm N}$ are arranged in increasing order.
Then, eq. (\ref{recursionk1}) becomes
\begin{eqnarray}
K_{\rm N+1} = h_{1,2}+2 U^*_{\rm N}\omega^*_{\rm N} U_{\rm N}^{*\dagger} -  U^{**}_{\rm N-1}\omega^{**}_{\rm N-1} U^{**\dagger}_{\rm N-1}.
\label{recursionk2}
\end{eqnarray}
Here, we introduce new unitary matrices
\begin{equation}
P^{*}_{\rm N}\equiv  U^{**\dagger}_{\rm N-1}  U^{*}_{\rm N}\quad{\rm and} \quad P_{\rm N}\equiv  U^{*\dagger}_{\rm N}  U^{}_{\rm N+1}.
\end{equation}
By using these matrices, we finally obtain the main recursion relation for the corner Hamiltonians
\begin{eqnarray}
\bar{K}_{\rm N+1} = P_{\rm N}^{*\dagger}  h_{1,2} P^*_{\rm N}+ 2 \omega^*_{\rm N} -  P_{\rm N}^{*\dagger}\omega^{**}_{\rm N-1} P^*_{\rm N},
\label{recursionrg1}
\end{eqnarray}
where 
\begin{equation}
\bar{K}_{\rm N+1}\equiv U^{*\dagger}_{\rm N+1}K_{\rm N+1} U^*_{\rm N+1}= P_{\rm N+1} \omega_{\rm N+1} P^\dagger_{\rm N+1}.
\end{equation}
The above expression is very useful for the NRG calculation, since the unitary matrix appearing in the right-hand side of eq. (\ref{recursionrg1}) is only $P^*_{\rm N}$.

\subsection{iterative diagonalization}

We solve the eigenvalue problem of the corner Hamiltonian for a sufficiently large N recursively.
We first set up the initial corner Hamiltonian for N=2 and 3 and diagonalize them.
We then obtain an extended corner Hamiltonian by the recursion relation (\ref{recursionrg1}), and  diagonalize $\bar{K}_{\rm N+1}$ numerically by the Householder method,
\begin{eqnarray}
\bar{K}_{\rm N+1} P_{\rm N+1} = P_{\rm N+1}\omega_{\rm N+1}.
\end{eqnarray}
By using these extended matrices, we can return to (\ref{recursionrg1}) with incrementing N$\to$N+1.

In the above recursive step, the dimension of $\bar{K}_{\rm N+1}$ is doubled.
Thus, we retain half of  the {\it lower-energy} eigenvalues, in each recursion step, and the corresponding eigenvectors $P_{\rm N+1}$,  after a sufficiently large N.
If the number of retained eigenvalues is $m$, the dimension of $\bar{K}_{\rm N+1}$ is $2m$. We then keep the lower-half eigenvalues, and $P_{\rm N+1}$ becomes a $2m\times m$ matrix, which plays the role of the projection matrix. 
In the context of retaining the lower-energy eigenvalues, the present algorithm is faithful to Wilson's original NRG method rather than the DMRG that keeps larger eigenvalues of the reduced density matrix .

\subsection{self-consistent  equations}

After a sufficient number of iterations,  the matrices will converge to those in the thermodynamic(${\rm N}\to\infty$) limit.
In this sense, we omit the subscript N when discussing the matrices in ${\rm N}\to\infty$.
Then we can write down a set of the self-consistent equations,
\begin{eqnarray}
\bar{K} &=&   P^{* \dagger}  h_{1,2} P^* + 2 \omega^* -  P^{*\dagger}\omega^{**} P^{*},
\label{selfconsistent}\\
\bar{K}P&=&P\omega,
\label{selfev} \\
P^\dagger P& =& I,
\label{selforth}
\end{eqnarray}
which determine the eigenvalue structure of the corner Hamiltonian in the thermodynamic limit.
Here, we note that the dimensions of the matrices are as follows:
\begin{eqnarray*}
&{}&\omega: m\times m, \qquad \omega^* : 2m\times 2m, \qquad \omega^{**}:4m\times 4m \\
&{}&P: 2m\times m, \qquad P^* : 4m\times 2m, \qquad \bar{K}:2m\times 2m.
\end{eqnarray*}  
Unfortunately, we cannot illustrate an impressive graphical expression for the above self-consistent equations, unlike those of the CTMs(see chap.13 in ref. \cite{Baxbook}). 
However, we think that it is important to write down the closed form of the self-consistent equations in considering the fixed point of the DMRG.

Further, we deduce self-consistent equations for the renormalized Hamiltonian, which may be  relevant to the DMRG; 
Once the renormalized corner Hamiltonian is obtained, the renormalized Hamiltonian is also reproduced from (\ref{recursionh}) as 
\begin{equation}
\bar{H}^*= \omega^* - P^{*\dagger} \omega^{**} P^*,
\end{equation}
where  $\bar{H}^* \equiv U^{*\dagger}H^* U^*$.
In connection with the DMRG,  an alternative  expression
\begin{equation}
\bar{H}=  P^{*\dagger} h_{1,2} P^* + \omega^* - P^{*\dagger} \omega^{**} P^*,
\label{renormalizedh}
\end{equation}
is more suitable, where $\bar{H} \equiv U^{*\dagger}H U^*$. 
Substituting (\ref{renormalizedh}) into (\ref{selfconsistent}), we obtain
\begin{equation}
\bar{K}=\bar{H}+\omega^*.
\end{equation}
Further substituting this equation of $\bar{K}$ into (\ref{selfev}),
we finally arrive at
\begin{equation}
\bar{H}P=P\omega - \omega^* P,
\end{equation}
or equivalently,
\begin{equation}
\sum_{s_1'}\bar{H}(s_1,s_1')P(s_1')=[P(s_1),\omega].
\label{hocommu}
\end{equation}
In the last equation, we recover the indices of the bare spins and thus (\ref{hocommu}) is the equation for the $m\times m$ matrices.

Equations (\ref{selforth}), (\ref{renormalizedh}) and (\ref{hocommu}), as a set, are also the equivalent self-consistent equations for the renormalized Hamiltonian in the thermodynamic limit. 
Here, it should be recalled that $\omega$ is the eigenvalue matrix not of the reduced density matrix but of the corner Hamiltonian. 
For the case of the integrable model, however,  we have $\omega\propto \ln\rho_{d}$ with the common eigenvectors, where $\rho_{d}$ is the eigenvalue matrix of the reduced density matrix in the DMRG with an appropriate normalization.

Of course the above-obtained self-consistent equations are too formal for practical use.
In order to solve them, we have to employ the iterative method described in the previous subsection.

\section{Results for the $S=1/2$ XXZ chain}

\begin{figure}[ht]
\begin{center}
\epsfig{file=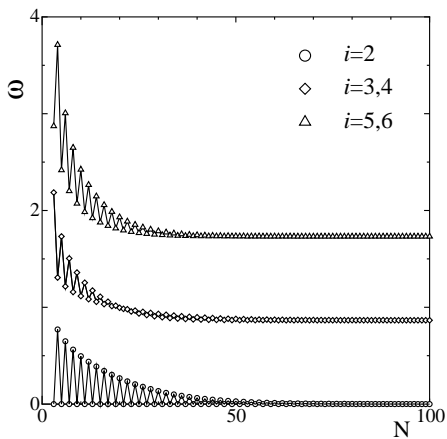,width=6cm}
\end{center}
\caption{Size dependence of lowest 7  eigenvalues of the corner Hamiltonian for the XXZ model with $\Delta=2$. The NRG calculation is performed with $m=200$.
The ground state($i=1$) corresponds to the horizontal axis. The $i=3,5$ and $i=5,6$ states are two-fold degenerating.}
\label{flow1}
\end{figure}

In this section, we demonstrate the NRG of the corner Hamiltonian for the $S=1/2$ XXZ chain.
We first discuss the Ising-like anisotropic case($\Delta>1$), for which  the corner Hamiltonian has a discrete eigenvalue structure in the thermodynamic limit.
The NRG calculation is performed with the retained number of bases $m=200$.
Figure \ref{flow1} shows the N dependence of the lowest seven  eigenvalues for $\Delta=2$. In the figures, the ``ground state'' energy is set to be zero.
For a small N, we can see the oscillation with respect to N=even or odd, for which the total $S^z$ of the system takes an integer or a half integer alternately.
As N increases, we can see that the eigenvalues rapidly converge to the constant values corresponding to the thermodynamic limit.

The exact eigenvalue structures of the corner Hamiltonian and the reduced density matrix of the XXZ model in the thermodynamic limit are well known for  $\Delta>1$, where the parameterization $\Delta\equiv \cosh \lambda$ is very useful.\cite{Baxbook,Bax1,Bax2}
Through the diagonal representation of the CTM, we can extract the exact spectrum of the corner Hamiltonian as
\begin{equation}
\omega=\frac{\sinh \lambda}{2}\cdot {\rm diag}(0,0,1,1,2,2,3,3,3,3\cdots ), 
\label{omega-diag}
\end{equation}
where diag($\cdots$) denotes  a diagonal matrix whose diagonal entry is $\cdots$.
Since the wavefunction of the XXZ chain is represented as $\Psi(\lambda)\propto A(\lambda-\mu)A(\mu)$ on the basis of the corresponding 6-vertex model, we also have the exact spectrum of the reduced density matrix
\begin{equation}
\rho_d(\lambda)\propto e^{-\alpha \omega},
\label{rho-diag}
\end{equation}
where $\rho_d$ is the eigenvalue matrix of the reduced density matrix.
The coefficient $\alpha$ can be identified as
\begin{equation}
\alpha=\frac{4\lambda}{\sinh \lambda},
\end{equation}
according to  Baxter's parametrization of the Boltzmann weight of the 6-vertex model.
Here, it should be noted that the above exact spectrum (\ref{rho-diag}) with (\ref{omega-diag}) was certainly reproduced by the DMRG, as in Refs.\cite{Kaulke,oha}.

\begin{figure}[ht]
\begin{center}
\epsfig{file=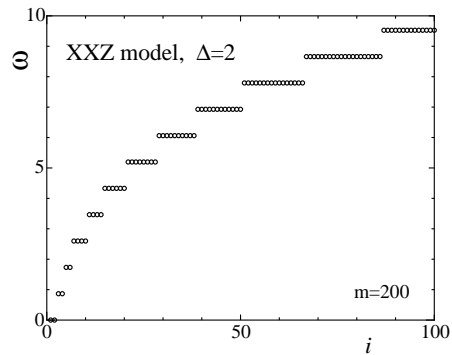,width=6cm}
\end{center}
\caption{Lowest 100 eigenvalues of the corner Hamiltonian for the XXZ model with $\Delta=2$ for ${\rm N}=200$. The NRG calculation is performed with $m=200$.}
\label{spectrumdelta2}
\end{figure}

Figure \ref{spectrumdelta2} shows the lowest 100 eigenvalues of the corner Hamiltonian for the XXZ model with $\Delta=2$, which are obtained after a sufficient number of iterations.
The correspondence between the results of the exact solution, the  corner Hamiltonian NRG  and the DMRG\cite{dmrgcal} is perfect within the level of numerical accuracy.
This implies that the NRG algorithm for the corner Hamiltonian  works successfully.

Next we discuss the XXZ model with $\Delta\le 1$, which has the gapless ground state. It is well known that  the corresponding CTM is not normalizable in the thermodynamic limit. 
In other words, $\omega \to 0$ as N increases. 
However, the size dependence of the spectrum of the CTM or the corner Hamiltonian can be  analyzed in the framework of the finite size scaling with the CFT;
the conformal mapping from the upper half-plane into the helical stairway geometry yields the spectrum of the CTM with an infinitesimal transfer angle, which can be associated with the corner Hamiltonian.\cite{peschel-troung,cardy,Kleban}
However, actual numerical calculations of the corner Hamiltonian for a large but finite size system have been difficult,  away from the free fermion model.

We define the energy gap of the corner Hamiltonian as $\Delta\omega \equiv \omega^{(1)} -\omega^{(0)}$ for ${\rm N}=$even, where the superscripts with the bracket denote a ``quantum number'' for the  energy level taking account of the degeneracy structure; e.g.,
$\Delta\omega$ corresponds to the energy difference between the first step and ground step of the stairwaylike spectrum.
Then, the CFT argument yields the size dependence of $\Delta\omega$ as
\begin{equation}
\Delta\omega \propto \frac{1}{\ln({\rm N}/c)},
\label{fsscft}
\end{equation}
where $c$ is an ultraviolet cutoff parameter.\cite{peschel,cardy}

\begin{figure}[ht]
\begin{center}
\epsfig{file=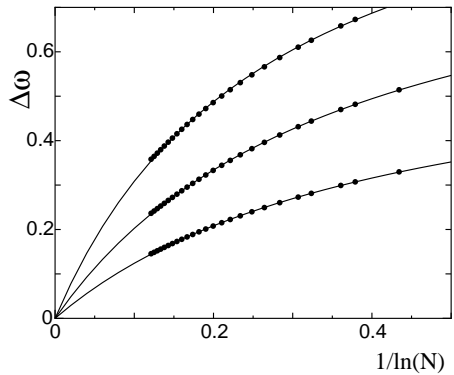,width=6cm}
\end{center}
\caption{Size dependence of gap $\Delta\omega$ for the XXZ chain in the gapless regime up to N$\simeq$4000. The solid symbols indicate the NRG results. The lines are the fitted curves with eq. (\ref{fitting}).
The data correspond to $\Delta=1.0$, $0.5$ and  $0.0$ from top to bottom.}
\label{fssgap}
\end{figure}

\begin{figure}[ht]
\begin{center}
\epsfig{file=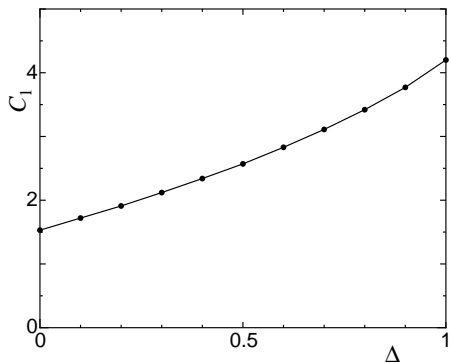,width=6cm}
\end{center}
\caption{Fitting result of  $C_1$ in (\ref{fitting}) for the XXZ chain of $0\le \Delta\le 1$. The line is a guide for eyes.}
\label{coef}
\end{figure}

In Fig. \ref{fssgap}, we show the size dependences of  $\Delta\omega$ for $\Delta=0.0$, 0.5 and 1.0, which are obtained with $m=1000$.  
We have confirmed that the $m$ dependence of $\Delta\omega$ is negligible up to ${\rm N}\sim 4000$.
In order to check (\ref{fsscft}), we examine a fitting of the form:
\begin{equation}
\Delta\omega = \frac{C_1}{\ln {\rm N} + C_2},
\label{fitting}
\end{equation}
where $C_1$ and $C_2$ are fitting parameters.
Here we note that  the parameter $C_2$ is definitely requested, since a variable in a logarithm should be dimensionless. 
In Fig. \ref{fssgap}, we can see that the NRG data is well fitted by (\ref{fitting}).
We show the obtained values of  $C_1$ in Fig. \ref{coef}.
The coefficient $C_1$ may be connected to the scaling dimension in CFT.
For this purpose, however, we need to carefully calculate the velocity with another numerical method.
A further study is clearly desired for the analysis of the coefficient $C_1$.

\section{Application to $S=1$ chains}

Let us examine the corner Hamiltonian NRG  for the $S=1$ Heisenberg spin chain, which is a typical example of nonintegrable models.
Since the $S=1$ Heisenberg chain is in the gapful phase,  the corner Hamiltonian and the reduced density matrix exhibit discrete structures in their eigenvalue spectra.
Indeed we have confirmed that,  as N increases, the low-energy eigenvalues rapidly converge to those in the thermodynamic limit.

In order to compare them  systematically, we define the logarithm of the eigenvalue spectrum of the reduced density matrix as 
\begin{equation}
\omega_D = - \frac{1}{\alpha} \ln \rho_d,
\end{equation}
where we assume that the largest eigenvalue of $\rho_d$ is normalized to be unity.
Here, we should recall that the direct relationship between the reduced density matrix and the corner Hamiltonian such as eq. (\ref{rho-k}) is not established for the non-integrable model, implying that the coefficient $\alpha$ is unknown a priori.
Thus, we set $\alpha\equiv -\ln \rho_d^{(1)}/\omega^{(1)}$; we scale $\omega_D$ so that its excitation gap corresponds to that of $\omega$, and then compare their stairwaylike structures.

\begin{figure}[ht]
\begin{center}
\epsfig{file=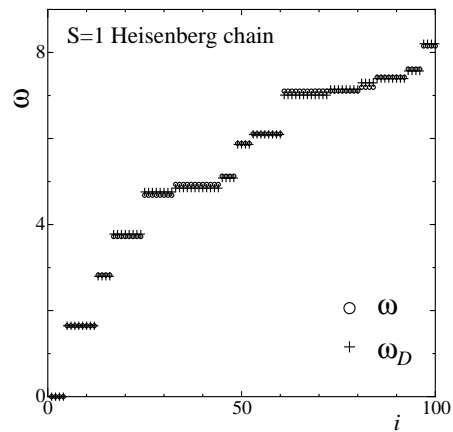,width=6cm}
\end{center}
\caption{Eigenvalue structure of the corner Hamiltonian for the $S=1$ Heisenberg model.}
\label{s1ev}
\end{figure}

In Fig. \ref{s1ev}, we show the comparison between the NRG result for the corner Hamiltonian and the $\omega_D$ spectrum obtained with the DMRG, for which we have used $\alpha=0.3942$. 
Both the NRG and DMRG calculations are performed with $m=200$. 
Since the integrability does not hold for the $S=1$ Heisenberg chain, we can see that the regular stairway structure is disturbed.
However, the remarkable point is that the dominant structures of the spectra for the corner Hamiltonian and the reduced density matrix are very similar;
Since $\omega_D$ is scaled to $\omega$ by one parameter $\alpha$,  
 the NRG for the corner Hamiltonian can contain almost the same level of information as the DMRG for the $S=1$ Heisenberg model.
In ref.\cite{oha}, we have found that the eigenvalue structure of the reduced density matrix seems to have universal asymptotic behavior not only for the integrable models but also for a class of 1D non-integrable models.
The above one-parameter scalability with respect to $\alpha$ may be related with such universal asymptotics of the eigenvalues.

In order to clarify the nature of the corner Hamiltonian further, we investigate the $S=1$ bilinear-biquadratic chain
\begin{equation}
h_{n,n+1}= \vec{S}_n\cdot\vec{S}_{n+1}+\beta(\vec{S}_n\cdot\vec{S}_{n+1})^2, 
\label{blbq}
\end{equation}
where $\vec{S}$ is the $S=1$ spin matrix.
This model has been intensively studied in the context of the Haldane gap.
$\beta=\pm 1$ is exactly solvable and the ground state is gapless.\cite{SU3,TB}
$\beta=1/3$ is the AKLT model whose ground state is exactly represented as the valence-bond-solid(VBS) state, which is a typical example of the matrix product form of the wavefunction.\cite{AKLT}
From the DMRG point of view, the AKLT model is particularly important, since the DMRG with $m=4$ can exactly reproduce the VBS state.\cite{m4}
Thus, it is interesting to know how the spectrum of the corner Hamiltonian for (\ref{blbq}) behaves, in contrast with the DMRG.

\begin{figure}[ht]
\begin{center}
\epsfig{file=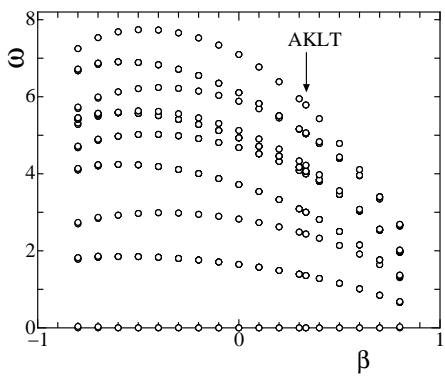,width=6cm}
\end{center}
\caption{$\beta$-dependence of lowest 64 eigenvalues of the corner Hamiltonian for the $S=1$ bilinear-biquadratic model.
An open circle corresponds to a step of the  almost degenerating eigenvalues in the stairwaylike eigenvalue structure.
The arrow indicates the AKLT point($\beta=1/3$).}
\label{betaflow}
\end{figure}

In Figure \ref{betaflow}, we show the $\beta$-dependence of the lowest 64 eigenvalues for $-0.8\le \beta \le 0.8$.
For $-0.5 \le \beta \le 0.8$, the NRG iteration sufficiently converges to the thermodynamic limit within $m=400$ and N=400.
As $\beta$ approaches $\pm1$,  the convergency becomes worse, since the exactly solvable points($\beta=\pm1$) are gapless.
Thus, we improve the accuracy up to $m=800$ and  N=2000 for $\beta\le -0.6$.
In the vicinity of the gapless points($\beta < -0.8 $  or $\beta > 0.8$), unfortunately,  we do not succeed in  extracting the spectrum of the thermodynamic limit.

In Fig. \ref{betaflow}, we see only a few points of levels for a fixed $\beta$, since the corner Hamiltonian has the almost degenerating eigenvalue structure illustrating the stairwaylike spectrum as in Fig. \ref{s1ev}, 
An interesting point is that the rearrangement of the spectrum clearly occurs, as $\beta$ varies from  $-0.8$ to $0.8$.
For instance, we can see that the 2nd and 3rd excitation levels merge as $\beta\to 1$.
In the vicinity of $\beta=\pm1$,  the eigenvalue spectrum is expected to have a nearly regular structure governed by the integrability of the massless fixed point  and the most relevant perturbation associated with $\beta$,
although the corner Hamiltonian at the massless fixed point itself is not normalizable.
The parameter $\beta$ connects these two integrable limits continuously.
Thus, the dominant structure of the spectrum is reconstructed adiabatically as $\beta$ varies between $-1< \beta <  1$.
The present result demonstrates such reconstruction of the structure of the spectrum.

Another important aspect of Fig. \ref{betaflow} is that $\beta=1/3$ is not any special point for the corner Hamiltonian.
Since the two-body Hamiltonian (\ref{blbq}) at $\beta=1/3$ is the projection operator for the $S=2$ state on the  nearest-neighboring bond,\cite{AKLT} the ``ground state'' of the corner Hamiltonian itself is identical to the VBS state for the AKLT Hamiltonian.
Accordingly, the convergence of the NRG iteration for $\beta=1/3$ is very fast; after a few iterations, the eigenvalue spectrum reaches the thermodynamic limit.
However, we find that the spectrum of the corner Hamiltonian behaves continuously around $\beta=1/3$;
the dimension of the corner Hamiltonian clearly increases with respect to the system size, while the eigenvalues of the reduced density matrix for the VBS state are exactly truncated by $m=4$.
This implies that the corner Hamiltonian  certainly involves information on the excitations(in the sense of the usual Hamiltonian), while the reduced density matrix is determined only by the ground-state wave function.
A further analysis of the deviation between these two spectra around $\beta=1/3$ may provide important information for understanding the correlation effects in the 1D quantum systems.

\section{Discussion}

In this paper, we have discussed the real-space renormalization group for the corner Hamiltonian.
We have derived the recursion relation for the corner Hamiltonian and its self-consistent equations in the thermodynamic limit, where the lower-energy states of the corner Hamiltonian are more relevant.
For the integrable model, we have verified that the corner Hamiltonian NRG yields the equivalent of the exact eigenvalue spectrum of the reduced density matrix.
Indeed we have shown that the exact spectrum is reproduced for the XXZ chain of $\Delta>1$.
For the massless case, the size dependence of the excitation gap of the corner Hamiltonian is consistent with the logarithmic behavior expected by the CFT, although a further study is desired for the direct connection of the fitting results with the scaling dimension.
For the $S=1$ Heisenberg chain, we have found that the dominant structure of $\omega_D$ can be scaled to $\omega$ by  one parameter $\alpha$.
Moreover,  the spectral flow for the bilinear-biquadratic chain  exhibits a clear reconstruction of the spectrum  between the two integrable points($\beta=\pm1$)

As far as the relation (\ref{rho-k}) is established,  the present results imply that we do not need the diagonalization of the super block Hamiltonian(in the DMRG sense).
This suggests that  a further simplification of the DMRG algorithm  might be possible in principle.
Of course,  we still have a gap to the practical use of the corner Hamiltonian NRG for calculating the physical quantities of the nonintegrable cases, where  the parameter $\alpha$ is not known a priori.
However, it is important that the corner Hamiltonian involves the same level of implications as the DMRG even for the non-integrable case.

From the theoretical viewpoint, we have seen that the renormalization group for the corner Hamiltonian provides a mount of interesting problems.
It is intriguing to know how we can use the self-consistent equations (\ref{selfconsistent}), (\ref{selfev}), (\ref{selforth}) or (\ref{selforth}), (\ref{renormalizedh}), (\ref{hocommu}) for an analysis of the matrices in the thermodynamic limit.
For the critical system,  the finite size analysis of the spectrum is explained by the CFT;
the direct estimation of the scaling dimension from the fitting data is a remaining problem.
For the nonintegrable case, we have seen the two aspects of the corner Hamiltonian:
the one-parameter scalability for the dominant structure of the spectra between the corner Hamiltonian and the reduced density matrix for the $S=1$ Heisenberg chain and, in contrast,  their deviation for the AKLT model($\beta=1/3$). 
In addition to the above, we should remark on the boundary condition of the corner Hamiltonian; 
Clearly the corner Hamiltonian treated here is only the right(or left) block in the DMRG and  the free boundary condition is imposed both for the center and edge spins. In contrast, the center spins in the DMRG are directly entangled with the other block of the system. 
This may be a key point in discussing the role of the entanglement of the states in the 1D quantum systems, where $\alpha$ may have an important physical meaning even for the nonintegrable case. 
We also note that the connection to the density-matrix spectrum in the finite size DMRG, which we have not treated here, is a remaining important problem from the practical point of view.

Our motivation at the start is to answer the question: what kind of low-energy effective theory is obtained by the DMRG in the Wilson's sense. 
We think that the present corner Hamiltonian approach provides a possible way to address the question and, at the same time, stimulates further investigations on the issue.

\acknowledgments

The author thanks T. Nishino for valuable discussions.
This work is partially supported by Grants-in-Aid for Scientific Research (B)(No.17340100), (C)(No.16540332) and (C)(No.17540317). It is also partially supported by a Grant-in-Aid for Scientific Research in Priority Areas A (No. 17038011).

%%%%%%%%%%%%%%%%%%%%%%%%%%%%%%%%%%


\begin{thebibliography}{99}

%\begin{references}


\bibitem{wilson} K.G. Wilson: Rev. Mod. Phys. {\bf 47},(1975) 773.

\bibitem{KWW} H.R. Krisina-murthy, J.M. Wilkins and K.G. Wilson: Phys. Rev. B {\bf 21}, (1980) 1003.

\bibitem{White} S.R. White: Phys. Rev. Lett. {\bf 69}, (1992) 2863; Phys. Rev. {\bf B 48}, (1993) 10345.
\bibitem{springer} ``Density-matrix renormalization''eds. I. Peschel, X. Wang, M. Kaulke and K. Hallberg, Springer (1998)

\bibitem{Scholl} U. Schollw\"ock:  Rev. Mod. Phys. {\bf 77}, (2005) 259.

\bibitem{or} S. Ostllund and Rommer:  Phys. Rev. Lett {\bf 75}, (1995) 3537.

\bibitem{Bax1}R. J. Baxter: J. Math. Phys.  {\bf 9}, (1968) 650.

\bibitem{Bax2}R. J. Baxter: J. Stat. Phys. {\bf 15}, (1976) 485;  J. Stat. Phys. {\bf 17}, (1977) 1.

\bibitem{Baxbook}R. J. Baxter: ``Exactly solved models in Statistical mechanics'', Academic Press 1982.

\bibitem{thacker} H.B. Thacker: Physica D {\bf 18}, (1986) 348.

\bibitem{ctmrg} T. Nishino and K. Okunishi: J. Phys. Soc. Jpn. {\bf 65}, (1996) 891; J. Phys. Soc. Jp. {\bf 66}, (1997) 3040. 



\bibitem{Kaulke} I. Peschel, M. Kaulke and \"O. Legeza: Ann. Physik. {\bf 8}, (1999) 153.

\bibitem{oha} K. Okunishi, Y. Hieida and Y. Akutsu: Phys. Rev. E {\bf 59},  (1999) R6227.

\bibitem{holzhey} C. Holzhey, F. Larsen and F. Wilczek: Nucl. Phys. B {\bf 424}, (1994) 443.

\bibitem{vidal} G. Vidal, J. Latorre, E. Rico and A. Kitaev: Phys. Rev. Lett. {\bf  90}, (2003) 227902.

\bibitem{Calabrese} P. Calabrese and J. Cardy: JSTAT P06002 (2004).

\bibitem{chung} M.-C. Chung and I. Peschel: Phys. Rev. B, {\bf 64}, (2001) 064412.

\bibitem{peschel} I. Peschel: JSTAT P06004  (2004).

\bibitem{peschel2} I. Peschel: JSTAT P12005 (2004).


\bibitem{dmrgcal} In the actual DMRG computation, the party symmetry is imposed; we use the reflection of the left block as the right block Hamiltonian.

\bibitem{peschel-troung} I. Peschel and T.T. Troung: Z. Phys. B {\bf 69}, (1987) 395.

\bibitem{cardy} J.L. Cardy and I. Peschel: Nucl. Phys. B {\bf 300}, (1988) 377.

\bibitem{Kleban} P. Kleban and I. Peschel: Z. Phys. B {\bf 101}, (1996) 447.

\bibitem{pearce} B. Davies and P.A. Pearce: J. Phys. A: Math. Gen. {\bf 23}, (1990) 1295.

%\bibitem{itoyama} H. Itoyama and H.B. Thacker, Nucl. Phys. B 


\bibitem{SU3}C.K. Lai: J. Math. Phys. {\bf 15}, (1974) 1675.; B. Sutherland:  Phys. Rev. B {\bf 12}, (1975) 3795.

\bibitem{TB}  L.A. Takhtajan: Phys. Lett. A  {\bf 87}, (1982) 479; H.M. Babujian: Nucl.Phys.B {\bf215}, (1983) 317.

\bibitem{AKLT} I. Affleck, T. Kennedy, E. H. Lieb, and H. Tasaki:  Phys. Rev. Lett. {\bf 59}, (1987) 799.


\bibitem{m4} The matrix dimension of the VBS state is $m=2$. However, if the reflection symmetry of the left and right blocks in the DMRG calculation is assumed, the dimension of the reduced density matrix is doubled due to the twofold degeneracy. 




%\end{references}

\end{thebibliography}
\end{document}